\documentclass{article}

\usepackage[preprint]{neurips_2024}

\usepackage[pdftex]{graphicx}
\usepackage[dvipsnames]{xcolor}
\usepackage[utf8]{inputenc} 
\usepackage[T1]{fontenc}    
\usepackage{hyperref}       
\usepackage{url}            
\usepackage{booktabs}       
\usepackage{amsfonts}       
\usepackage{nicefrac}       
\usepackage{microtype}      
\usepackage{xcolor}         
\usepackage{float}
\usepackage[raster,most]{tcolorbox}
\usepackage{natbib}
\setcitestyle{numbers,square}

\title{Reinforcement Learning Outperforms Supervised Fine-Tuning: A Case Study on Audio Question Answering}

\definecolor{gray}{HTML}{777777}
\definecolor{}{HTML}{00008B}

\author{
Gang Li$^\dag$, Jizhong Liu$^\dag$, Heinrich Dinkel, Yadong Niu, Junbo Zhang, Jian Luan  \\
\\
Xiaomi Corporation, China\\
\\
\texttt{\{ligang5,liujizhong1\}@xiaomi.com}\\
}

\begin{document}

\maketitle

\renewcommand{\thefootnote}{}
\footnotetext{$^\dag$ Equal contribution.}
\setcounter{footnote}{0}

\begin{abstract}
Recently, reinforcement learning (RL) has been shown to greatly enhance the reasoning capabilities of large language models (LLMs), and RL-based approaches have been progressively applied to visual multimodal tasks. However, the audio modality has largely been overlooked in these developments. Thus, we conduct a series of RL explorations in audio understanding and reasoning, specifically focusing on the audio question answering (AQA) task. We leverage the group relative policy optimization (GRPO) algorithm to Qwen2-Audio-7B-Instruct, and our experiments demonstrated state-of-the-art performance on the MMAU \textit{Test-mini} benchmark, achieving an accuracy rate of 64.5\%. The main findings in this technical report are as follows: 1) The GRPO algorithm can be effectively applied to large audio language models (LALMs), even when the model has only 8.2B parameters; 2) With only 38k post-training samples, RL significantly outperforms supervised fine-tuning (SFT), indicating that RL-based approaches can be effective without large datasets; 3) The explicit reasoning process has not shown significant benefits for AQA tasks, and how to efficiently utilize \textit{deep thinking} remains an open question for further research; 4) LALMs still lag far behind humans auditory-language reasoning, suggesting that the RL-based approaches warrant further explorations.
Our project is available at \url{https://github.com/xiaomi-research/r1-aqa} and \url{https://huggingface.co/mispeech/r1-aqa}.
\end{abstract}

\section{Introduction}

The latest breakthroughs in large language models (LLMs) have greatly enhanced their reasoning abilities, particularly in mathematics and coding. DeepSeek-R1~\cite{deepseekr1}, a pioneering innovator, has demonstrated how reinforcement learning (RL) can effectively improve LLMs' complex reasoning capabilities. Although chains-of-thought (CoT) is a simple rule-based reward model, it effectively aids reasoning tasks by simulating the human thought process. Therefore, reinforcement learning is likely to achieve performance far surpassing supervised fine-tuning (SFT) through simple methods. Recently, many researchers ~\cite{chen2025r1v, ahamomentvisual} have attempted to incorporate simple yet ingenious RL methods into visual modality understanding and reasoning tasks. 

However, the audio modality has largely been overlooked in recent developments. Although large audio language models (LALMs) have been increasingly proposed, such as Qwen2-Audio~\cite{Qwen2-Audio} and Audio Flamingo 2~\cite{Audio-Flamingo}. LALMs still rely on pre-trained modules along with SFT to construct systems. In fact, it is not that RL-based approaches are unsuitable for LALMs; rather, tasks such as automatic speech recognition (ASR) and automated audio captioning (AAC) are simple descriptive tasks~\cite{MMAU}. More complex logical reasoning tasks are needed to fully explore the potential of RL in the audio modality. 

Audio Question Answering (AQA) is a multimodal task that involves understanding and reasoning based on audio content to generate accurate responses to questions. It integrates both the auditory modality and linguistic modality, making it particularly suitable for evaluating complex logical reasoning capabilities. It demands the ability to extract meaningful insights from raw audio signals, infer implicit relationships, and provide contextually relevant answers. Due to its inherent complexity, AQA serves as an ideal benchmark for testing the effectiveness of reinforcement learning approaches. In addition, AQA can be considered as an advanced technology built upon automated audio captioning.

Based on the above reasons, we take AQA as a topic to explore the effectiveness of RL and \textit{deep thinking} in the audio modality. Meanwhile, it is encouraging to see that some researchers have already started some attempts~\cite{Audio-Reasoner, Audio-CoT}. In this report, we present a successful technical attempt, where the group relative policy optimization (GRPO) algorithm~\cite{DeepSeekMath}, with a small-scale dataset, improves the reasoning performance of the AQA task via Qwen2-Audio-7B-Instruct~\cite{Qwen2-Audio}. Our experiments demonstrate state-of-the-art performance on the MMAU test-mini benchmark, achieving an accuracy rate of 64.5\%. In summary, The main findings are as follows:
\begin{itemize}
    \item The GRPO algorithm can be directly and effectively applied to the audio modality and LALMs, even to Qwen2-Audio-7B-Instruct with only 8.2 billion parameters.
    \item With only 38k post-training samples, RL outperforms supervised fine-tuning (SFT), indicating that RL-based approaches can be effective without large datasets. In other words, the application of RL does not entirely depend on data scale.
    \item The explicit reasoning process has not shown significant benefits for AQA tasks, and how to efficiently leverage \textit{deep thinking} or step-by-step reasoning remains an open question for further research.
    \item Based on test results from the MMAU \textit{test-mini} benchmark, LALMs still lag far behind humans auditory-language reasoning, suggesting that the RL-based approaches warrant further explorations.
\end{itemize}

\section{Related Works}

\textbf{Audio Question Answering.} AQA is a multimodal task that involves understanding and reasoning over audio content to generate accurate responses to questions. In LLMs' frameworks, AQA  builds upon AAC. While AAC focuses on generating descriptive textual captions for audio, AQA requires a deeper comprehension of complex acoustic patterns, temporal relationships, and contextual information embedded in the audio. Although researchers have achieved good performance on the AAC task~\cite{rank2, EnCLAP, rank3,SLAM-AAC, rank4, loae}, AQA remains a multimodal challenge, which combines auditory and linguistic modalities, making it ideal for evaluating complex reasoning. can be categorized based on audio type into single-audio and multi-audio tasks, and based on response format into selection-based and open-ended questions. As illustrated in Figure~\ref{fig:audio_question}, we focus on the most common single-audio task with selection-based answers. Additionally, a multiple-choice setting with a single correct answer presents a significant generation-verification gap~\cite{roadsleadlikelihoodvalue}, making it a suitable setting for evaluating the effectiveness of RL in the audio modality. RL tends to perform well when verification is easy, but generation is complex.

\begin{figure}[t]
\centering
\begin{tcolorbox}[title=\text{\textbf{Q:} Based on the given audio, what event is associated with the clickety-clack sounds?}, label=QA, colbacktitle=Black, colback=White]
\textbf{A.} Train passing over tracks \\
\textbf{B.} Footsteps on a wooden floor \\
\textbf{C.} A machine operating in a factory \\
\textbf{D.} A horse galloping on a road
\end{tcolorbox}
\caption{An example of a question and choices based on audio.}
\label{fig:audio_question}
\end{figure}

\textbf{Multimodal Reasoning.} Recent studies have indicated that \textit{deep thinking} can enhance the reasoning performance of LLMs~\cite{deepseekr1,openai2024learning}. DeepSeek-R1 and OpenAI-o1 have significantly improved the reasoning capabilities, particularly in multi-step tasks such as coding and mathematics, sparking renewed interest in RL and CoT. Furthermore, RL and CoT are playing an increasing role in multimodal large language models. For instance, VisualThinker R1 Zero~\cite{ahamomentvisual} implements RL in visual reasoning on a 2B non-sft model. LLaVA-CoT~\cite{llava-cot} outperforms its base model by 7.4\% using only 100k post-training samples and a simple yet effective inference time scaling method. Recent studies have also explored the CoT-based method in the audio modality. Audio-CoT~\cite{Audio-CoT} shows some improvements via zero-shot CoT, but the improvement is very limited. Audio-Reasoner~\cite{Audio-Reasoner} has achieved significant improvements through extensive fine-tuning data and a complex reasoning process. But it lacks thorough ablation studies, and some of these processes may be redundant. Overall, studies on \textit{deep thinking} in the audio modality is still limited.

\textbf{Large Audio Language Models.} LALMs can generally be divided into two categories: audio understanding and audio generation. This study focuses on the field of audio understanding. There are already many representative models, such as Qwen2-Audio~\cite{Qwen2-Audio},  Audio Flamingo 2~\cite{Audio-Flamingo}, and SALMONN~\cite{salmonn}. However, they are all SFT-based models, and whether RL can unlock their potential remains to be studied.

\section{Method}
\label{sec:method}
In this section, we present the training method for our exploration, which leads to the state-of-the-art performance on the MMAU \textit{Test-mini} benchmark. The goal is to apply the GRPO algorithm directly into to LALMs. 

\begin{figure}[t]
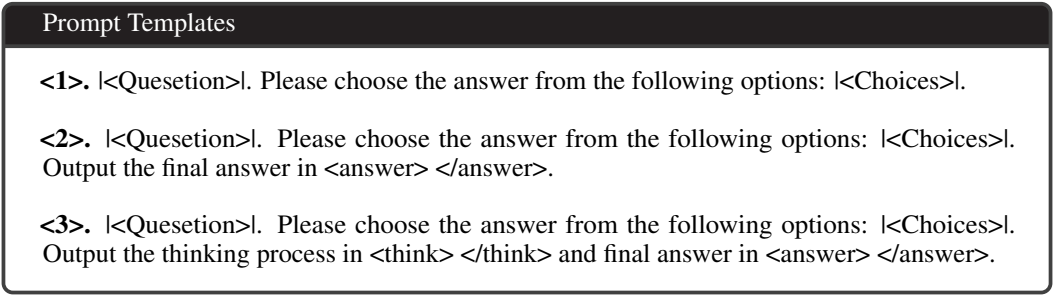

\centering
\begin{tcolorbox}[title=Prompt Templates, label=prompt_template, colbacktitle=Black, colback=White]
\textbf{<1>.} |<Quesetion>|. Please choose the answer from the following options: |<Choices>|. \\ \\
\textbf{<2>.} |<Quesetion>|. Please choose the answer from the following options: |<Choices>|. Output the final answer in <answer> </answer>. \\ \\
\textbf{<3>.} |<Quesetion>|. Please choose the answer from the following options: |<Choices>|. Output the thinking process in <think> </think> and final answer in <answer> </answer>.
\end{tcolorbox}
\caption{Different instruction prompts for reinforcement learning and supervised fine-tuning, where <1> is used for supervised fine-tuning, while both <2> and <3> are used for reinforcement learning. Audio-based questions and options are enclosed in |<Question>| and |<Choices>|.}
\label{fig:prompt}
\end{figure}

Our exploration is based on the Qwen2-Audio-7B-Instruct model~\cite{Qwen2-Audio}. We leverage the GRPO algorithm along with a customized chat template and prompting strategy to enhance its reasoning capabilities. Compared to SFT, RL may be a more efficient and effective way to adapt to downstream tasks~\cite{ahamomentvisual, llava-cot}. In SFT, instruction tuning requires a large amount of training data. Whether this condition is necessary in RL is one of the key questions. We train our models with the instruction template in Figure~\ref{fig:prompt}. For each question in the dataset, the model generates a response in the <answer> </answer> template, then it is optimized using a RL objective. The difference is that Prompt <2> does not require the model to explicitly output its reasoning process, whereas Prompt <3> does (i.e., CoT). As the <think> </think> cannot be directly supervised and the <answer> </answer> serves merely as a template, SFT adopts the simplest Prompt <1>. 

A review of the GRPO algorithm used for training is provided, mainly referring~\cite{ahamomentvisual, DeepSeekMath}. To ease the burden of training an additional value function approximation model in proximal policy optimization (PPO)~\cite{ppo}, GRPO employs the average reward of sampled response from the policy model as the baseline in computing the advantage. Specifically, given an input question $q$, a group of responses $\{o_1, o_2, \cdots, o_G\}$ is first sample, and their corresponding rewards corresponding rewards $\{r_1, r_2, \cdots, r_G\}$ are computed using the reward model. The advantage is subsequently computed as:
\begin{equation}
    \hat{A}_{i, t} = \widetilde{r}_i = \frac{r_i- {\rm mean}(\mathbf{r})}{{\rm std}(\mathbf{r})}
\end{equation}
The policy model is subsequently optimized by maximizing the Kullback-Leibler objective:
\begin{equation}
\begin{split}
    \mathcal{J}_{GRPO}(\theta) &= \mathbb{E}_{q \sim P(Q), \{o_i\}_{i=1}^G \sim \pi_{\theta_{old}}(O|q)}  
    \Bigg[ \frac{1}{G} \sum_{i=1}^G \frac{1}{|o_i|} \sum_{t=1}^{|o_i|} \Bigg\{ \min \Bigg[
    \frac{\pi_\theta(o_{i,t} | q, o_{i,<t})}{\pi_{\theta_{old}}(o_{i,t} | q, o_{i,<t})} \hat{A}_{i,t}, \\
    & \quad \text{clip} \left( \frac{\pi_\theta(o_{i,t} | q, o_{i,<t})}{\pi_{\theta_{old}}(o_{i,t} | q, o_{i,<t})}, 1 - \epsilon, 1 + \epsilon \right)  \hat{A}_{i,t}  
    \Bigg] - \beta \mathbb{D}_{KL}\left[\pi_{\theta} || \pi_{ref}\right] \Bigg\} \Bigg]
\end{split}
\label{eq:GRPO-obj}
\end{equation}
where $\pi_{\theta}$ and $\pi_{old}$ are the current and former policy, and $\epsilon$ and $\beta$ are hyper-parameters introduced in PPO. Responses are evaluated by a rule-based reward function in terms of their format and correctness:

\begin{itemize}
    \item If the response provides a correct final answer, the model obtains an accuracy reward of +1.
    \item If the response encloses the thinking in <think> </think> (if it is in the format) and the final answer in <answer> </answer>, the model obtains a format reward of +1.
    \item otherwise, the model receives 0 reward.
\end{itemize}

\section{Experiments}
In this study, we train our models via full fine-tuning, low-rank adaptation (LoRA)~\cite{lora}, and reinforcement learning. To effectively evaluate generalization, the evaluation follows an out-of-distribution testing approach, where the training and test sets come from different data sources.

\subsection{Experimental Setup}
\renewcommand{\thefootnote}{1}
\textbf{Datasets.} The training data is from the AVQA dataset~\cite{AVQA}, which is designed to audio-visual question answering by providing a comprehensive resource for understanding multimodal information in real-life video scenarios. It comprises 57015 videos depicting daily audio-visual activities, accompanied by 57335 specially designed question-answer pairs.  The questions are crafted to involve various relationships between objects and activities. We only use audio-text pairs of the training subset and change the ``video'' in the question to ``audio''. Our training set has approximately 38k samples\footnotemark. The test data is from the MMAU dataset~\cite{MMAU}. MMAU is a comprehensive dataset designed to evaluate advanced LALMs on complex tasks requiring expert-level knowledge and reasoning. It consists of 10000 audio clips, each paired with human-annotated natural language questions and answers, encompassing domains such as speech, environmental sounds, and music. MMAU released a full suite comprising 1000 \textit{Test-mini} samples, and the last 9000 \textit{Test} samples questions are available without their answers. Thus we only use the \textit{Test-mini} subset as the test set.

\footnotetext{The AVQA training set originally consists of approximately 40k samples. However, only about 38k samples are used, as some data sources have become invalid.}

\textbf{Implementation Details.} The RL models are trained using eight NVIDIA H800 GPUs, with each device running a batch size of 1. The model is trained for 500 steps with a learning rate of $1 \times 10^{-6}$ and a temperature of 1.0. All hyper-parameters are given in Table~\ref{tab:exp_settings}. We conduct comparative experiments on full fine-tuning and LoRA, with each device running a batch size of 4 and a gradient accumulation step of 1. The SFT models are trained using the AdamW optimizerr with a learning rate of $5 \times 10^{-6}$ for a total of 4 epochs, with a checkpoint saved every 200 steps. The optimal iteration result is selected for final analysis.

\begin{table}[t]
    \centering
        \caption{Hyper-parameters of reinforcement learning with the GRPO algorithm.}
        \resizebox{0.5\columnwidth}{!}{
    \begin{tabular}{l c}
             \hline
            \toprule
        \textbf{Setting} & \textbf{Value} \\
        \hline
        Batch Size per Device & 1 \\
        Gradient Accumulation Steps & 2 \\ 
        Training Steps & 500 \\
        Learning Rate & $1 \times 10^{-6}$ \\
        Temperature & 1.0 \\
        Maximum Response Length & 512 \\
        Number of Responses per GRPO Step & 8 \\
        Kullback-Leible Coefficient & 0.04 \\
        \hline
    \end{tabular}}
    \label{tab:exp_settings}
\end{table}

\subsection{Main Results}
To evaluate the effectiveness of reinforcement learning, we compare both SFT methods and RL methods on MMAU \textit{Test-mini} benchmark. Specifically, There are three types of strategies: direct inference by LALMs, fine-tuning LALMs with SFT, and fine-tuning LALMs with RL. The main results are given in Table~\ref{tab:main_results}. The baseline is derived from the top six entries on the official MMAU leaderboard and two recent studies based on reinforcement learning. Using the GRPO algorithm and Prompt <2>, we achieved state-of-the-art average accuracy on the MMAU \textit{Test-mini} benchmark. Nevertheless, all LALMs still lag far behind humans. This suggests that there is still a pressing need to improve the understanding and reasoning capabilities of LALMs.

\begin{table}[t]
    \renewcommand{\arraystretch}{1.3}
    \small
    \centering
    \caption{Accuracies (\%) on MMAU \textit{Test-mini} benchmark. \textit{Human} denotes the result of human testing, and \textit{Strong Cap.} denotes that the model inputs are strong audio captions generated by Qwen2-Audio-Instruct. Blue represents methods based on reinforcement learning, and an underline indicates the method with the highest average accuracy.}
        \begin{tabular}{cccccc}
             \hline
            \toprule
             \multicolumn{1}{c}{\textbf{Model}} & \multicolumn{1}{c}{\textbf{Method}} & \multicolumn{4}{c}{\textbf{MMAU Test-mini}} \\
             \midrule
              & & Sound & Music & Speech & Average  \\
              \hline
              \multicolumn{6}{l}{\textit{\textbf{Baselines:}}} \\
                \color{gray}{-----} & \color{gray}{Human\textsuperscript{*}} & \color{gray}{86.31} & \color{gray}{78.22} & \color{gray}{82.17} & \color{gray}{82.23} \\
                Gemini Pro 2.0 Flash & Direct Inference\textsuperscript{*} & 56.46 & 58.68 & 51.65 & 55.60 \\
                Audio Flamingo 2 & Direct Inference\textsuperscript{*} & 61.56 & \textbf{73.95} & 30.93 & 55.48 \\
                GPT4o + Strong Cap. & Direct Inference\textsuperscript{*} & 57.35 & 49.70 & \textbf{64.86} & 57.30 \\
                Llama-3-8B-Instruct + Strong Cap. & Direct Inference\textsuperscript{*} & 50.75 & 48.93 & 55.25 & 52.10 \\
                Gemini Pro v1.5 & Direct Inference\textsuperscript{*} & 56.75 & 49.40 & 58.55 & 54.90 \\
                Qwen2-Audio-7B-Instruct & Direct Inference\textsuperscript{*} & 54.95 & 50.98 & 42.04 & 49.20 \\
                Qwen2-Audio-7B-Instruct & CoTA~\cite{Audio-Reasoner} & 60.06 & 64.30 & 60.70 & 61.71 \\
                Qwen2-Audio-7B-Instruct & Zero-Shot-CoT~\cite{Audio-CoT} & 61.86 & 56.29 & 55.26 & 57.80 \\                
                \hline
                \multicolumn{6}{l}{\textit{\textbf{Ours:}}} \\
                Qwen2-Audio-7B-Instruct & Full + Prompt <1> & 60.96 & 49.19 & 45.35 & 51.80  \\
                Qwen2-Audio-7B-Instruct & LoRA + Prompt <1>  & 58.26 & 58.08 & 52.58 & 56.40  \\
                Qwen2-Audio-7B-Instruct & \color{blue}{\underline{GRPO + Prompt <2>}} & \textbf{69.37} & 66.77 & 57.36 & \textbf{64.50} \\
                Qwen2-Audio-7B-Instruct & \color{blue}{GRPO + Prompt <3>} & 66.67 & 62.87 & 53.75 & 61.10  \\
             \hline
        \end{tabular}
    \label{tab:main_results}
    \begin{flushleft}
        \scriptsize{\textsuperscript{*} The data are sourced from the MMAU official website: \url{https://sakshi113.github.io/mmau_homepage/}}
    \end{flushleft}
\end{table}

\begin{figure}[t]
    \centering
    \includegraphics[width=0.48\linewidth]{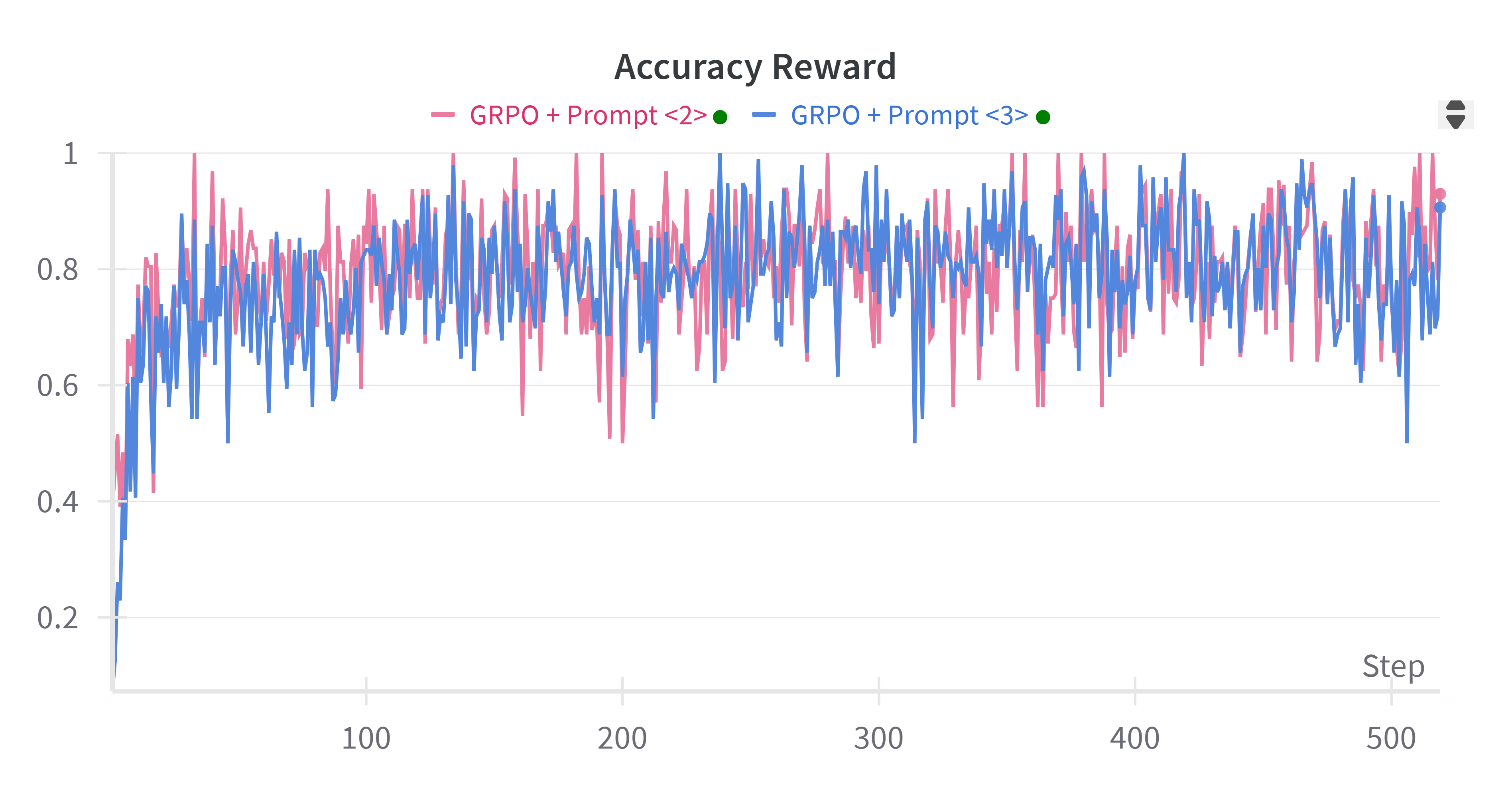} \hspace{0.02\linewidth}
    \includegraphics[width=0.48\linewidth]{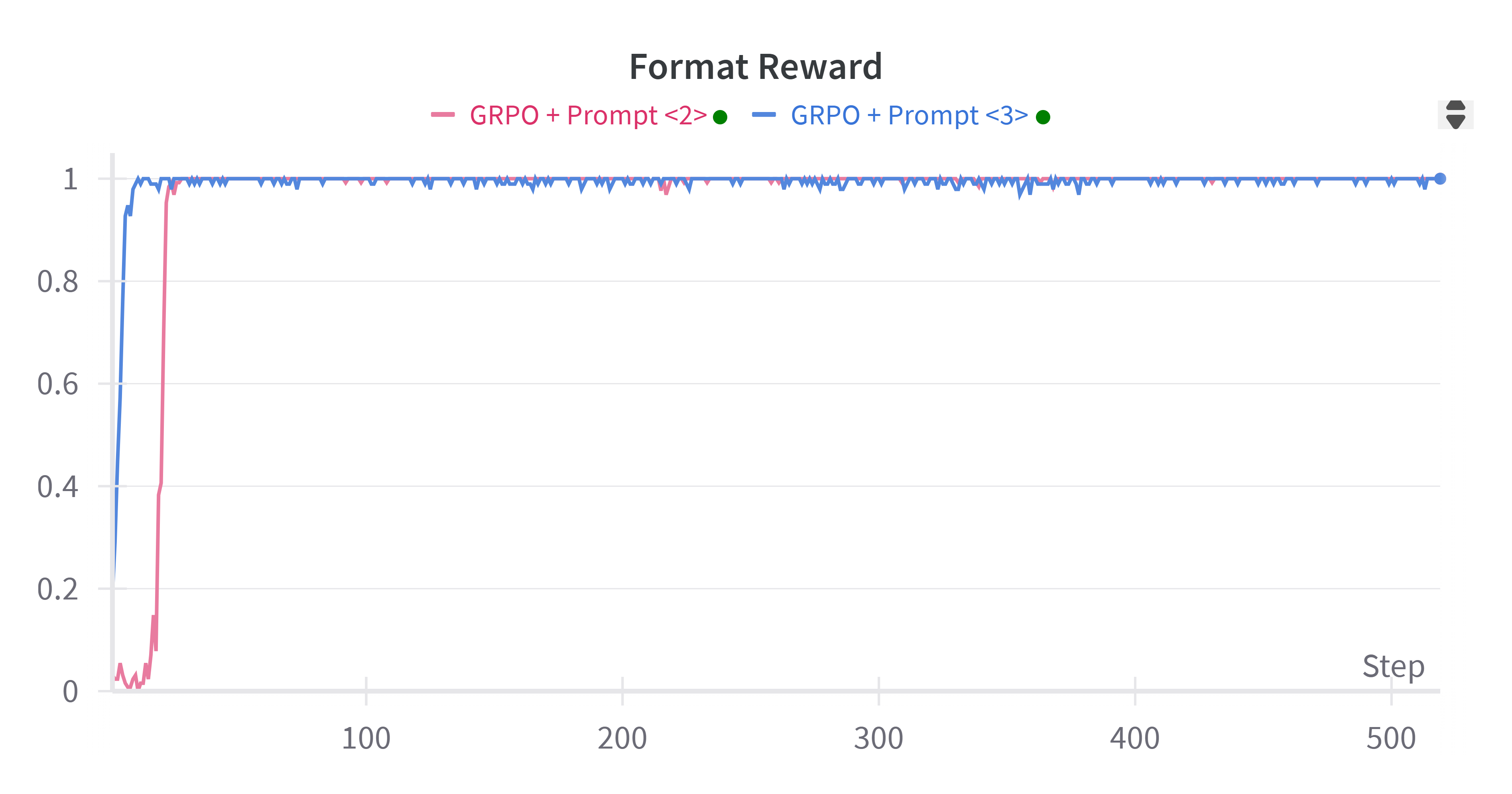}
    \\ 
    \includegraphics[width=0.48\linewidth]{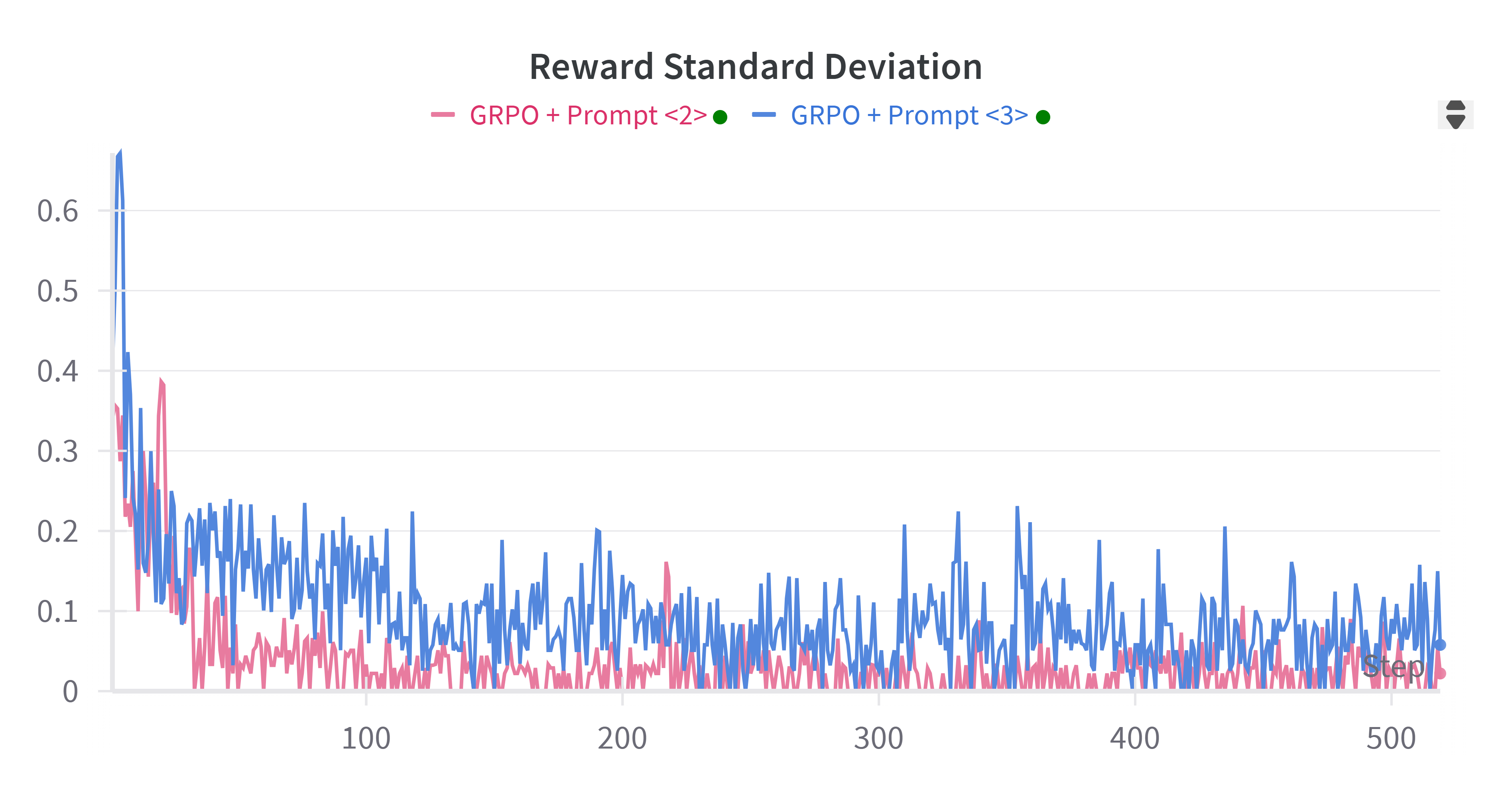} \hspace{0.02\linewidth}
    \includegraphics[width=0.48\linewidth]{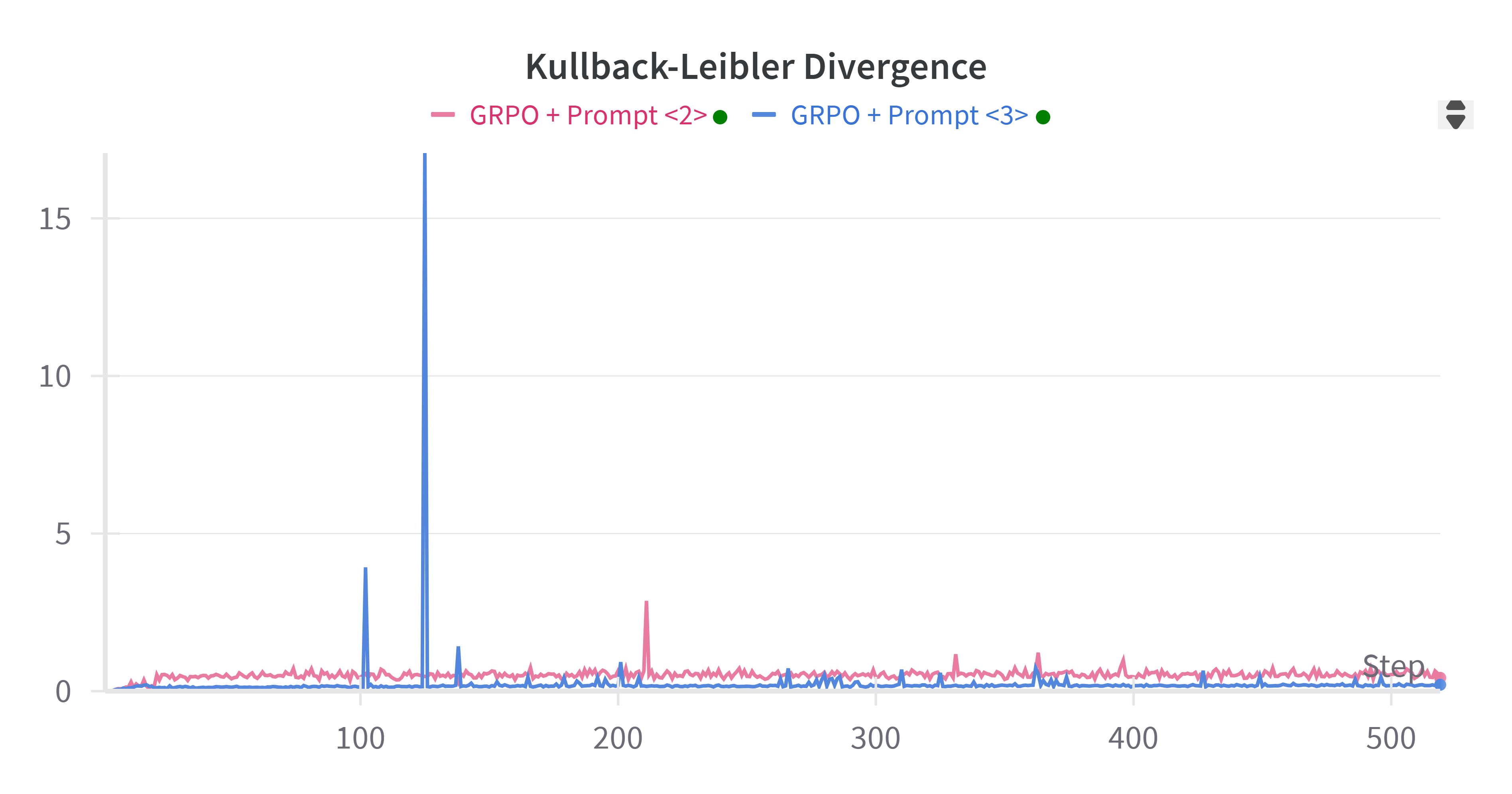}
    \\
    \includegraphics[width=0.48\linewidth]{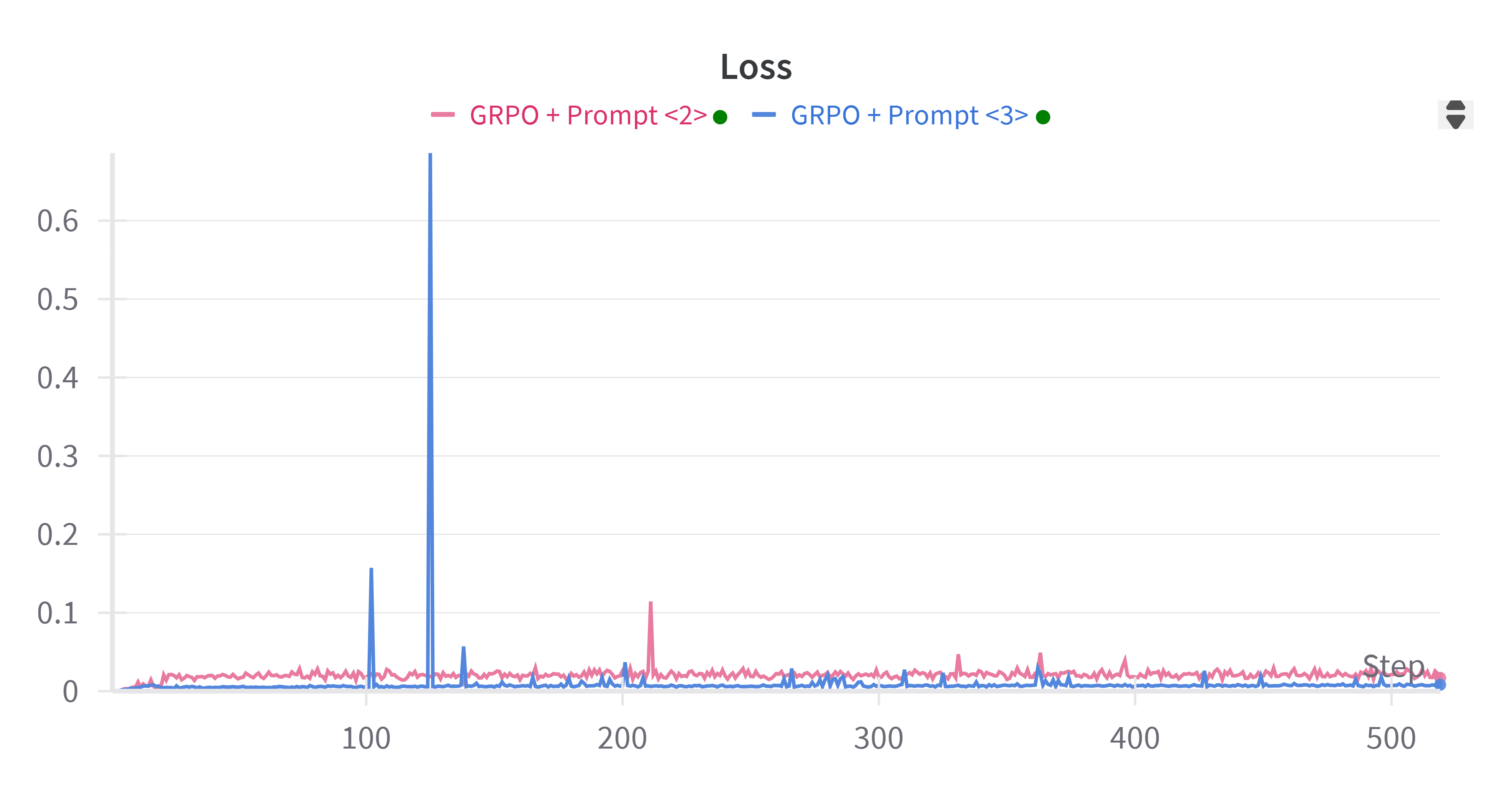} \hspace{0.02\linewidth}
    \includegraphics[width=0.48\linewidth]{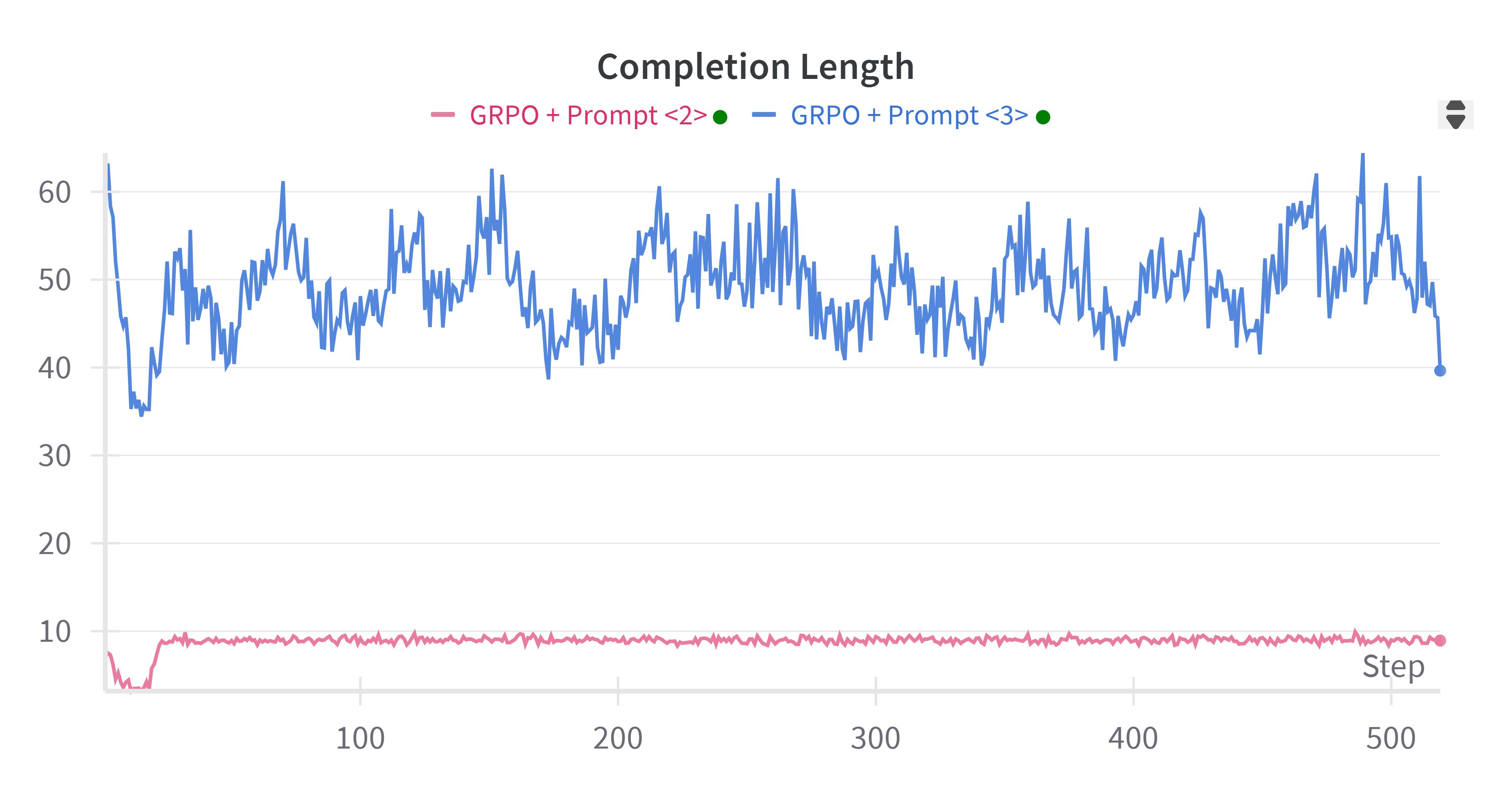}
    \caption{Convergence processes of the GRPO algorithm.}
    \label{fig:RL}
\end{figure}

Furthermore, \textit{deep thinking} methods demonstrate overall superior performance to classic SFT. The top four methods in Table~\ref{tab:main_results} are all RL or CoT approaches: GRPO + Prompt <2>, GRPO + Prompt <3>, Audio-Reasoner (CoTA)~\cite{Audio-Reasoner}, and Audio-CoT (Zero-Shot-CoT)~\cite{Audio-CoT}. AQA is a task that is challenging to generate but easy to verify. Extracting the correct answer from the audio content is highly challenging, while verifying options is straightforward. The AQA task empirically verifies the conclusion in ~\cite{roadsleadlikelihoodvalue} that tasks with a generation-verification gap are suitable for RL. Additionally, RL has demonstrated strong generalization with a limited amount of training data. Figure~\ref{fig:RL} illustrates the convergence process of the GRPO algorithm. Using only 38k AVQA training samples, we achieved state-of-the-art performance on out-of-distribution tests. This aligns with LLaVA-CoT~\cite{llava-cot} finding that structured thinking is well-suited for small-sample training. Whether the extensive and complex fine-tuning data utilized in Audio-Reasoner is truly necessary remains an open question. Another intriguing question is whether insights from the vision modality~\cite{ahamomentvisual} can be utilized to facilitate auditory reasoning in a non-SFT model.

However, we find some differences in the implementation of RL compared to previous studies~\cite{deepseekr1,ahamomentvisual}. Explicit CoT may not outperform letting the model think on its own. Neither the CoT templates in~\cite{Audio-Reasoner, Audio-CoT} nor the simple way <think> </think> in Prompt <3> outperforms directly prompting the model to generate <answer> </answer>. At least for AQA tasks, How \textit{deep thinking} and step-by-step reasoning contribute to the task requires further exploration.

\begin{figure}[t]
    \centering
    \includegraphics[width=0.48\linewidth]{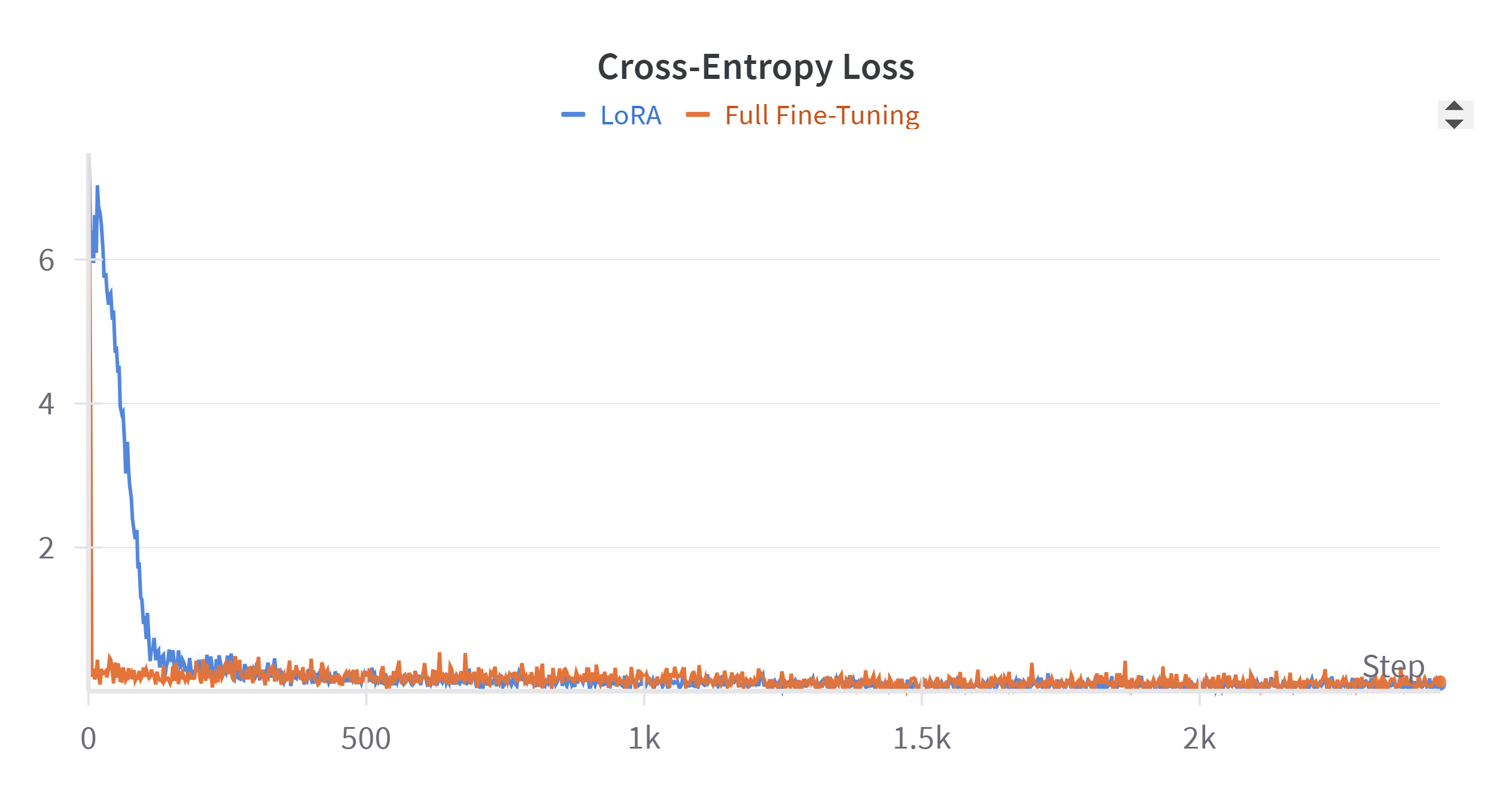} \hspace{0.02\linewidth}
    \includegraphics[width=0.48\linewidth]{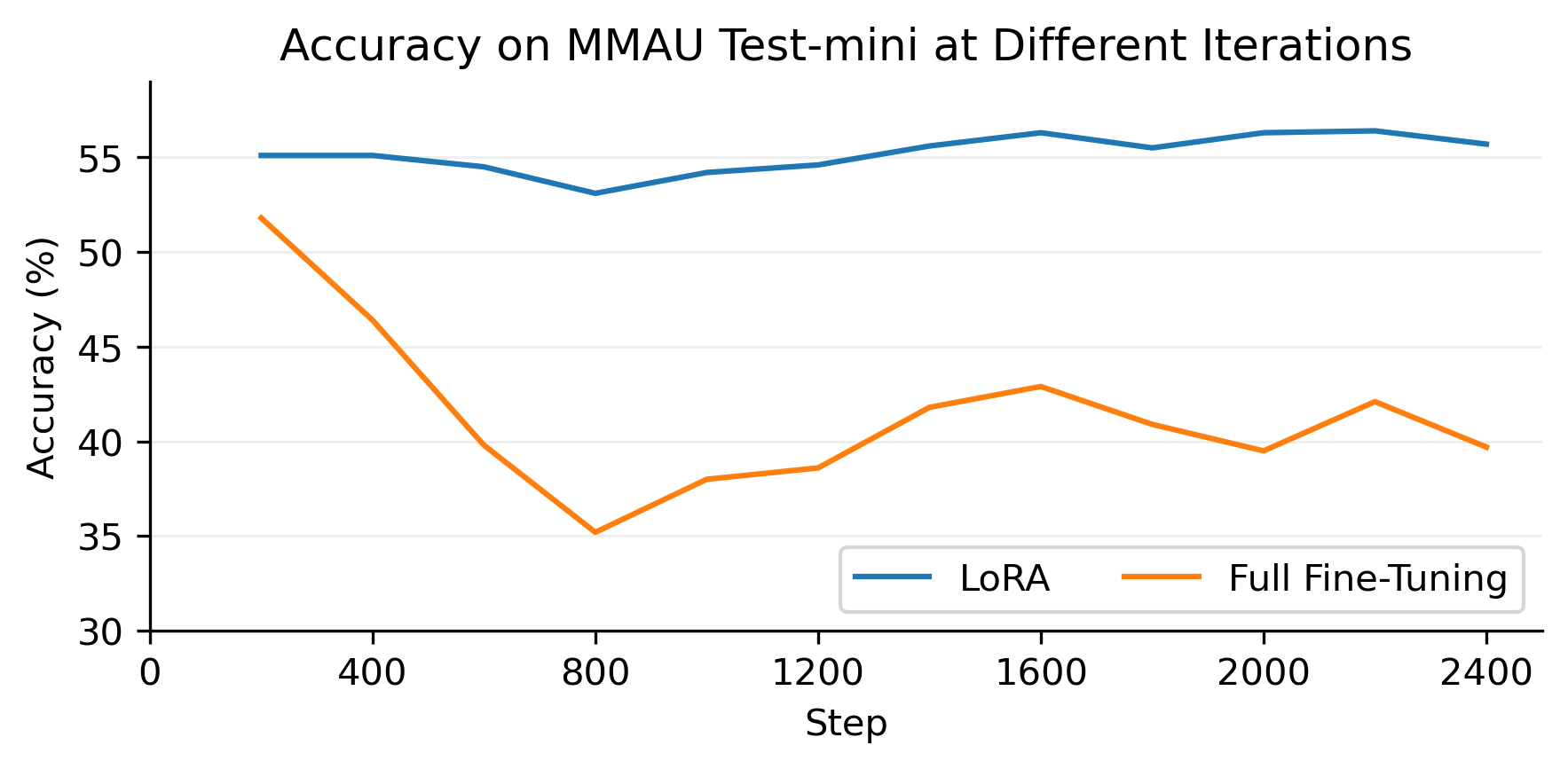}
    \caption{Convergence processes of the full and LoRA fine-tuning.}
    \label{fig:SFT}
\end{figure}

As shown in Figure~\ref{fig:SFT}, SFT struggles to converge with a small-scale dataset. Full fine-tuning appears to fit the training set quickly, but its accuracy on the out-of-distribution MMAU \textit{Test-mini} set decreases as the model fits the training set more closely. This further confirms that RL fine-tuning is more suitable than SFT for small-scale datasets. Although LoRA partially addresses the challenges of fine-tuning on small-scale datasets, SFT still performs worse compared to RL. The accuracy of LoRA + Prompt <1> is 8.1\% lower than that of GRPO + Prompt <2>, emphasizing the need to explore RL-based fine-tuning approaches.

\section{Conclusion}
This report presents a case study on reinforcement learning and supervised fine-tuning for audio question answering. We directly apply the GRPO algorithm to Qwen2-Audio-7B-Instruct, and achieve state-of-the-art performance on the MMAU~\textit{Test-mini} benchmark, with an accuracy rate of 64.5\%. This result indicates that reinforcement learning can be effectively applied to LALMs and audio multimodal, especially to AQA, which involves a significant generation-verification gap. Even if the model has a small number of parameters and the fine-tuning dataset is limited, it does not affect the implementation of reinforcement learning. In fact, on small-scale datasets, the strong generalization capability of reinforcement learning can be demonstrated even more clearly. However, the explicit reasoning process has not shown significant benefits for AQA tasks, and how to efficiently leverage \textit{deep thinking} and step-by-step reasoning remains an open question for further research. LALMs still lag far behind humans auditory-language reasoning, suggesting that reinforcement learning warrants further explorations. Our future research will focus on the effective integration of the chain-of-thought paradigm into the audio modality.


\begin{thebibliography}{1}

\bibitem{deepseekr1}
DeepSeek-AI, ``{DeepSeek-R1: Incentivizing Reasoning Capability in LLMs via
  Reinforcement Learning},'' \emph{arXiv preprint arXiv:2501.12948}, 2025.

\bibitem{chen2025r1v}
L.~Chen, L.~Li, H.~Zhao, Y.~Song, and Vinci, ``{R1-V}: Reinforcing super
  generalization ability in vision-language models with less than \$3,'' 2025,
  accessed: 2025-02-02. [Online]. Available:
  \url{https://github.com/Deep-Agent/R1-V}

\bibitem{ahamomentvisual}
H.~Zhou, X.~Li, R.~Wang, M.~Cheng, T.~Zhou, and C.-J. Hsieh, ``{R1-Zero's "Aha
  Moment" in Visual Reasoning on a 2B Non-SFT Model},'' \emph{arXiv preprint
  arXiv:2503.05132}, 2025.

\bibitem{llava-cot}
G.~Xu, P.~Jin, H.~Li, Y.~Song, L.~Sun, and L.~Yuan, ``{LLaVA-CoT}: Let vision
  language models reason step-by-step,'' \emph{arXiv preprint
  arXiv:2411.10440}, 2025.

\bibitem{Qwen2-Audio}
Y.~Chu, J.~Xu, Q.~Yang, H.~Wei, X.~Wei, Z.~Guo, Y.~Leng, Y.~Lv, J.~He, J.~Lin,
  C.~Zhou, and J.~Zhou, ``{Qwen2-Audio} technical report,'' \emph{arXiv
  preprint arXiv:2407.10759}, 2024.

\bibitem{Audio-Flamingo}
S.~Ghosh, Z.~Kong, S.~Kumar, S.~Sakshi, J.~Kim, W.~Ping, R.~Valle, D.~Manocha,
  and B.~Catanzaro, ``{Audio Flamingo 2}: An audio-language model with
  long-audio understanding and expert reasoning abilities,'' 2025.

\bibitem{MMAU}
S.~Sakshi, U.~Tyagi, S.~Kumar, A.~Seth, R.~Selvakumar, O.~Nieto, R.~Duraiswami,
  S.~Ghosh, and D.~Manocha, ``{MMAU: A Massive Multi-Task Audio Understanding
  and Reasoning Benchmark},'' \emph{arXiv preprint arXiv:2410.19168}, 2024.

\bibitem{Audio-Reasoner}
Z.~Xie, M.~Lin, Z.~Liu, P.~Wu, S.~Yan, and C.~Miao, ``Audio-reasoner: Improving
  reasoning capability in large audio language models,'' \emph{arXiv preprint
  arXiv:2503.02318}, 2025.

\bibitem{Audio-CoT}
Z.~Ma, Z.~Chen, Y.~Wang, E.~S. Chng, and X.~Chen, ``{Audio-CoT}: Exploring
  chain-of-thought reasoning in large audio language model,'' \emph{arXiv
  preprint arXiv:2501.07246}, 2025.

\bibitem{DeepSeekMath}
Z.~Shao, P.~Wang, Q.~Zhu, R.~Xu, J.~Song, X.~Bi, H.~Zhang, M.~Zhang, Y.~K. Li,
  Y.~Wu, and D.~Guo, ``Deepseekmath: Pushing the limits of mathematical
  reasoning in open language models,'' \emph{2402.03300}, 2024.

\bibitem{rank2}
J.~Kim, J.~Jung, M.~Jeon, S.~H. Woo, and J.~Lee, ``Expanding on {EnCLAP} with
  auxiliary retrieval model for automated audio captioning,'' DCASE2024
  Challenge, Tech. Rep. 108, May 2024.

\bibitem{EnCLAP}
J.~Kim, M.~Jeon, J.~Jung, S.~H. Woo, and J.~Lee, ``{EnCLAP++: Analyzing the
  EnCLAP Framework for Optimizing Automated Audio Captioning Performance},''
  \emph{arXiv preprint arXiv:2409.01201}, 2024.

\bibitem{rank3}
W.~Chen, X.~Li, Z.~Ma, Y.~Liang, A.~Jiang, Z.~Zheng, Y.~Qian, P.~Fan, W.-Q.
  Zhang, C.~Lu, J.~Liu, and X.~Chen, ``{SJTU-THU Automated Audio Captioning
  System for DCASE 2024},'' DCASE2024 Challenge, Tech. Rep.~2, May 2024.

\bibitem{SLAM-AAC}
W.~Chen, Z.~Ma, X.~Li, X.~Xu, Y.~Liang, Z.~Zheng, K.~Yu, and X.~Chen,
  ``Slam-aac: Enhancing audio captioning with paraphrasing augmentation and
  clap-refine through llms,'' in \emph{ICASSP 2025 - 2025 IEEE International
  Conference on Acoustics, Speech and Signal Processing (ICASSP)}, 2025, pp.
  1--5.

\bibitem{rank4}
J.~Liu, G.~Li, C.~Liu, J.~Zhang, H.~Dinkel, Y.~Wang, Z.~Yan, Y.~Wang, and
  B.~Wang, ``Leveraging {CED} encoder and large language models for automated
  audio captioning,'' DCASE2024 Challenge, Tech. Rep.~42, May 2024.

\bibitem{loae}
J.~Liu, G.~Li, J.~Zhang, H.~Dinkel, Y.~Wang, Z.~Yan, Y.~Wang, and B.~Wang,
  ``Enhancing automated audio captioning via large language models with
  optimized audio encoding,'' in \emph{Interspeech 2024}, 2024, pp. 1135--1139.

\bibitem{roadsleadlikelihoodvalue}
G.~Swamy, S.~Choudhury, W.~Sun, Z.~S. Wu, and J.~A. Bagnell, ``All roads lead
  to likelihood: The value of reinforcement learning in fine-tuning,''
  \emph{arXiv preprint arXiv:2503.01067}, 2025.

\bibitem{openai2024learning}
{OpenAI}, ``Learning to reason with llms,'' 2024. [Online]. Available:
  \url{https://openai.com/index/learning-to-reason-with-llms/}

\bibitem{salmonn}
C.~Tang, W.~Yu, G.~Sun, X.~Chen, T.~Tan, W.~Li, L.~Lu, Z.~Ma, and C.~Zhang,
  ``SALMONN: Towards generic hearing abilities for large language models,''
  \emph{arXiv preprint arXiv:2310.13289}, 2023.

\bibitem{ppo}
J.~Schulman, F.~Wolski, P.~Dhariwal, A.~Radford, and O.~Klimov, ``Proximal
  policy optimization algorithms,'' \emph{arXiv preprint arXiv:1707.06347},
  2017.

\bibitem{lora}
E.~J. Hu, Y.~Shen, P.~Wallis, Z.~Allen-Zhu, Y.~Li, S.~Wang, L.~Wang, and
  W.~Chen, ``{LoRA}: Low-rank adaptation of large language models,''
  \emph{arXiv preprint arXiv:2106.09685}, 2021.

\bibitem{AVQA}
P.~Yang, X.~Wang, X.~Duan, H.~Chen, R.~Hou, C.~Jin, and W.~Zhu, ``{AVQA}: A
  dataset for audio-visual question answering on videos,'' in \emph{Proceedings
  of the 30th ACM International Conference on Multimedia}, ser. MM '22, 2022,
  p. 3480–3491.

\end{thebibliography}
\end{document}